 \documentstyle[twocolumn,aps,eqsecnum,epsf]{revtex}
\begin{document}
\draft
\title{
Impurity in a Luttinger liquid away from half-filling: a numerical study
}
 \author{ Shaojin Qin$^a$, Michele Fabrizio$^b$, Lu Yu$^{a,c}$ }
 \address{$^a$ International Center for Theoretical Physics, 
 P.O. Box 586, 34100 Trieste, Italy.}
 \address{$^b$ International School for Advanced Studies, Via Beirut 
 2-4, I-34014 Trieste, Italy, and Istituto Nazionale di Fisica della 
 Materia, INFM.}
 \address{$^c$ Institute of Theoretical Physics, Academia Sinica,
 Beijing 100080, China.}
 \author{ Masaki Oshikawa$^d$, and Ian Affleck$^{d,e}$ }
 \address{$^d$Department of Physics and Astronomy, University of 
 British Columbia, Vancouver BC V6T 1Z1 Canada}
 \address{$^e$Canadian Institute for Advanced Research,
 University of British Columbia, Vancouver BC V6T 1Z1 Canada}
\date{\today}
\maketitle
\begin{abstract}
Conformal field theory gives quite detailed predictions for the low 
energy spectrum and scaling exponents of a massless Luttinger liquid 
at generic filling in the presence of an impurity.  While these 
predictions were verified for half-filled systems, there was till now 
no analysis away from this particular filling. Here, we fill in this 
gap by numerically investigating a quarter-filled system using the 
density matrix renormalization group technique.  Our results confirm 
conformal field theory predictions, and suggest that they are indeed 
valid for arbitrary fillings.  
\end{abstract}



\section{Introduction}

The behavior of a one-dimensional (1D) interacting Fermi system 
(Luttinger liquid) in the presence of a single impurity has been the 
subject of an intensive theoretical and numerical investigation in 
recent years, for its interesting anomalies with respect to higher 
dimensions, and its implications to a variety of physical 
problems as for instance the behavior of quantum wires (see e.g. 
Refs.\onlinecite{Sasha,Exp}) or the tunneling through a constriction 
in the fractional quantum Hall regime\cite{Moon}.  
Specifically, if the interaction is repulsive, the electrons at low 
energies feel the impurity potential as if it were effectively 
infinite.  On the contrary, for attractive interactions, the effective 
scattering close to the Fermi energy is vanishingly small 
(see e.g. Ref.\onlinecite{K&F}).  In their seminal paper, Kane and 
Fisher\cite{K&F} argued that the low energy behavior for 
repulsive interaction in fact corresponds to a chain disconnected at 
the impurity site.
In the renormalization group language of boundary conformal field
theory (CFT) (see e.g. Refs.\onlinecite{A&E,A&L}), for repulsive 
interaction the open boundary condition (OBC) is the stable fixed 
point while the periodic boundary condition (PBC) is unstable, and 
oppositely for attractive interaction (except for the presence of 
nontrivial fixed points for spinning electrons with spin anisotropic 
interaction\cite{K&Fbis,furu1}). 
CFT provides the low energy spectrum and the scaling exponents both 
for the approach to the appropriate stable fixed point and for the 
departure from the unstable one.    

For a half-filled Luttinger liquid, these predictions have been 
numerically verified\cite{A&E,Q&F}.  In the present work, we study 
numerically the non-half-filled system, which is the generic 
situation for realistic systems.  Compared with the half-filled case, 
the non-half-filled one has non-zero forward scattering phase shift 
at the Fermi surface, which makes the analysis more cumbersome.  
In spite of that, we show that, also away from half-filling, the 
correct fixed point boundary conditions are those predicted in 
Refs.\onlinecite{K&F,A&E,A&L}.

At the end we will discuss in detail  the boundary conditions in 
relation to some previous studies\cite{Fink} which gave results 
different from those mentioned above.

\section{Numerical calculation}

We study numerically the $S=1/2$ XXZ Heisenberg chains with anisotropy 
$a=J_z/J_{xy}=\pm 0.5$ at magnetization per site $M=1/4$, which 
correspond to quarter-filled Luttinger liquids with repulsive ($a=0.5$) 
or attractive interaction ($a=-0.5$).  We model the impurity by 
modifying the strength of a particular bond, scaling it by a factor 
$b$.  Specifically, we consider the following Hamiltonian
\begin{equation}
\begin{array}{lcl}
H & = & {\displaystyle \sum_{i=1}^{L-1}}
                \left[(S_{i}^+S_{i+1}^-+S_{i}^-S_{i+1}^+)/2
                             +a S_{i}^zS_{i+1}^z \right] \\
  &   &  \mbox{} + b \left[(S_{1}^+S_{L}^-+S_{1}^-S_{L}^+)/2
                             +a S_{1}^zS_{L}^z \right], 
\end{array}
\label{old1}
\end{equation}
where $S^+_i$ and $S^-_i$ are spin raising and lowering operators at 
site $i$, $S^z_i$ being its z-component. $L$ is the chain length.
$b=0$ and $b=1$ correspond to what we will define as {\sl ideal} OBC
and PBC , respectively.  We use the density matrix renormalization 
group (DMRG) method\cite{white} to calculate the low energy spectrum 
and overlaps between ground state wave functions.  To ensure the 
correct magnetization $M=1/4$, we insert four sites at each step of 
DMRG iteration.  We run DMRG for $a=\pm 0.5$, $L=8,12,\ldots,60$, and 
$b=0.0,0.05,0.1,\ldots,0.9,0.95,1.0$.  
The optimal number of states which are kept is 300 and the truncation 
errors are less than $5\times 10^{-7}$.  Like our previous DMRG 
calculation for magnetization $M=0$\cite{Q&F}, we calculate the lowest 
energy states with parity even and odd (with respect to the modified 
bond), and $z$ component of total spin $S_z^{tot}=ML$ and $ML+1$ for 
$M=1/4$.  We denote these energies in increasing order as $E_0$, $E_1$, 
$E_2$, and $E_3$.  For states with magnetization $M$, $E_0$ corresponds 
to the ground state energy while $E_2$ corresponds to the energy of the 
lowest excited state with different parity.  $E_1$ and $E_3$ correspond 
to states with $S_z^{tot}=ML+1$, but with different parity.  
In addition to these low energy levels, we calculate also the ground 
state wave function overlaps between PBC chains ($b=1$) and impurity 
chains ($b\not = 1$).

\section{Energy spectrum and scaling predicted in CFT}

The spin chain at non-zero M and with a modified  bond can be written 
in terms of a boson field, $\phi$, which is defined periodically:
$ \phi \equiv \phi + 2\pi R$, and of its conjugate momentum $\Pi$.  
A detailed derivation is given in Ref. 
\onlinecite{A&E} for the case $M=0$.  Here we just point out the 
differences that occur for non-zero $M$.
The compactification radius, $R$, and the spin-wave velocity, $v_s$,
depend both on $a$ and  on  M. Their values for $a=\pm 0.5$ and $M=1/4$
are given in Table I. The effective Hamiltonian is: 
\begin{equation}
{\cal H}  = \int dx {1 \over 2}\left[\Pi^2+
\left({d\phi \over dx}\right)^2\right] 
         -\delta (x)\left[V_f {d\phi \over dx}+
         V_b \cos \left({\phi \over  R}\right)\right].
\label {Ham}
\end{equation} 
[We will generally set the spin-wave velocity $v_s$ to $1$ in most of 
what follows.] 
Here the modified bond is parameterized in terms of two boundary 
operators, with coupling constants $V_f$ and $V_b$.
The $V_f$ term is allowed, and therefore expected to 
occur, whenever $S^z\to -S^z$ symmetry as well as translational
invariance are broken, which is the case for $M\neq 0$ and with a 
modified bond.  Specifically, $V_f$ and $V_b$ correspond to forward and
backward scattering in the Luttinger liquid, respectively.  Forward 
scattering is marginal but backward scattering is relevant (irrelevant) 
for $R>1/\sqrt{4\pi}$  ($R<1/\sqrt{4\pi}$).  

In cases where backscattering is irrelevant, which we define 
generically as PBC cases (not to be confused with ideal PBC with 
$b=1$), $V_f$ induces a discontinuity in $\phi$ at the impurity site, 
$\phi(0+) - \phi(0-) \equiv \phi(0)-\phi(L)=2R\delta$ modulus $2\pi R$. 
Therefore the mode expansion can be written as: 
\begin{equation}
\phi (x,t) = 2\pi R(n-{\delta \over \pi}){x\over L}
		+{{mt} \over {RL}}+\ldots  
\end{equation}  
Here the dots represent the harmonic oscillator modes. 
$n =S_z^{\hbox{tot}}-ML$ and $m$ are integer quantum numbers. 
(The $\phi$ periodicity is preserved in time by the discrete quantization of the 
momentum conjugate.)  Thus the finite-size 
spectrum is: 
\begin{eqnarray}
 E & = & e_0L+e_1+n\mu + {2\pi v_s\over L} \left[-{1\over 12} 
	+\pi R^2\left(n-\frac{\delta}{\pi} \right)^2 \right.\nonumber\\
   &   & \mbox{}\left. +{m^2\over 4\pi R^2}
                 + \sum_{n=1}^\infty n(m_n^L+m_n^R)\right].
\label{fss1}
\end{eqnarray}
$m_n^{L,R}$ are occupation numbers for bosonic excitations of momentum 
$\pm 2\pi n/L$.  The chemical potential, $\mu$, is related to the 
external magnetic field necessary to induce the magnetization, $M$, in 
the spin chain. $\delta$ is the analogous to a forward scattering phase 
shift.
Here we have included the zero point energy $-2\pi v_s/(12L)$ which was 
ignored in Ref. \onlinecite{A&E}. This formula differs slightly from 
the one in Ref. \onlinecite{A&L} because in that case spinless fermions 
were considered rather than $S=1/2$ spins.  While the models are 
identical under the Jordan-Wigner transformation, PBC on the spins {\it 
do not} correspond to PBC on the fermions.  

When the backscattering term is relevant, $\phi (0)$ gets pinned. In 
analogy with the previous case, we define this situation as OBC, again 
not to be confused with the ideal OBC ($b=0$).  Without forward 
scattering (for $M=0$) $\phi (0)$ will simply be pinned at a unique 
value, $\phi(0)=0$ assuming $V_b>0$.  With both 
backward and forward scattering present, we expect $\phi$ to be pinned 
at equal and opposite values, $\pm \phi_0$ at $x\to 0^{\pm}$, with a 
discontinuity at $x=0$.  
Pinning forces $m=0$ in the mode expansion, changes the wavefunctions 
of the harmonic modes from $e^{\pm i2\pi nx/L}$ to $\sin (\pi nx/L)$  
and reduces the zero point energy by a factor of 1/4.   Thus
\begin{eqnarray} 
E & = & e_0L+e_1+n\mu + {\pi v_s\over L}\left[-{1\over 24}+
	2\pi R^2\left(n-{\delta\over \pi} \right)^2  \right. \nonumber\\
  &   &\mbox{} \left. + \sum_{n=1}^\infty n m_n\right].
\label{fss2}
\end{eqnarray}
The parameters $e_0$, $v_s$, $R$ and $\mu$ being bulk properties are
the same for PBC and OBC cases.  The parameters $e_1$ and $\delta$, on 
the other hand, depend on the boundary conditions, as well as the 
$1/12$ and $1/24$ terms. The latter can be 
calculated by elementary methods for the XX spin chain with PBC or 
OBC, respectively.  They have also been calculated, using the Bethe 
ansatz for the general XXZ model.  They follow more generally from 
conformal invariance.\cite{Cardy1,Affleck}  
Note that forward scattering always introduces a discontinuity in 
$\phi$ at the origin, with or without pinning of $\phi$ depending on 
whether or not backscattering is relevant.  However, we expect the 
value of the discontinuity $\propto \delta$ to depend on whether or 
not pinning occurs.

The effect of a parity transformation on the various eigenstates were
worked out in Ref. \onlinecite{A&E}.  In the periodic case it takes
$m\to -m$, $m_n^L\leftrightarrow m_N^R$ and multiplies wavefunctions
by $(-1)^{n+m}$.  In the case of OBC it multiples wavefunctions by 
$(-1)^n$ and transform a bosonic operator $a_m$, corresponding to an 
excitation with wave function $\sin(\pi m x/L)$, into $(-1)^m a_m$.  

We are now in position to work out the quantum numbers and energies of 
the four lowest energy states.  Consider first 
the states with $S_z^{tot}=ML$.  These states have $n=0$.  The ground 
state has all other quantum numbers $0$ as well.  For PBC, a state of 
reversed parity can be obtained either by taking $m=1$ (and all other 
quantum numbers 0) in Eq. (\ref{fss1}) or by taking an odd parity 
linear combination of the states with $(m_1^L,m_1^R)=(1,0)$ or $(0,1)$ 
(and all other quantum numbers 0).  These two states have energies 
measured from the ground state energy of $(2\pi v_s/L)/(4\pi R^2)$ and  
$(2\pi v_s/L)$ respectively.  Thus which of these states has lower 
energy depends on whether $4\pi R^2>1$ or $4\pi R^2<1$.  We see from 
Table I that the first one has lower energy for $a=1/2$, while the 
second one has lower energy for $a=-1/2$.  
The states with $S_z^{tot}=ML+1$ have $n=1$.  The state with all other
quantum numbers 0 has reversed parity relative to the grounds state. 
For PBC the state with the same parity as the ground state is obtained 
again by choosing $m=1$ (for $4\pi R^2>1$) or either $m_1^L$ or 
$m_1^R=1$ (for $4\pi R^2<1$).  
For OBC, the lowest energy states for given $S_z^{tot}$ 
(i.e. $n$) are obtained by choosing all the $m_n=0$, while the parity 
reversed lowest energy states are obtained by choosing $m_1=1$.
We define new quantities, $d_i$, by writing the $1/L$ term in the energy as
\begin{equation} 
E_i = ...+{2\pi v_s\over L}\left( -{1\over 12}+d_i\right),
\label{fss} 
\end{equation} 
for PBC or OBC.  The values of the $d_i$ for the first four states for 
ideal PBC and OBC for $a=\pm 0.5$ are given in Table II.

The renormalization group flow in the vicinity of the PBC and OBC fixed 
points are determined by the permitted boundary operator, $\cal O$, 
of lowest scaling dimension, $x$.  Depending 
on whether this dimension obeys $x<1$ or $x>1$, the fixed point will be
unstable or stable.  We may calculate the (size dependent) corrections 
to the energies, in the vicinity of the fixed point, by doing lowest 
order perturbation theory in this operator.\cite{Cardy2}  
The dimensions of a boundary operator at the PBC and OBC 
fixed points can be read off from the finite size energies. 
Specifically, if a boundary operator modifies 
some quantum numbers, its dimension is simply given by the
term inside the square brackets of Eq. (\ref{fss1}) for PBC and
(\ref{fss2}) for OBC with the appropriate change of quantum numbers, 
minus the analogous term with the ground state quantum numbers. 

For instance, the leading boundary operator 
which can be defined at the PBC fixed point is the backscattering 
potential ${\cal O} = \cos (\phi /R)$ [see the Hamiltonian (\ref{Ham})].  
Since it changes $m$ by $\pm 1$, its scaling dimension is simply 
$x=1/4\pi R^2$. The lowest order perturbative calculation is quite 
different depending on whether $4\pi R^2>1$ or $<1$.  Let us denote 
$|0>$ and  $|2>$ the states corresponding to the energy levels $E_0$ 
and $E_2$ previously introduced, and use $d_0$ and $d_2$ defined in 
Eq.(\ref{fss}).  If $4\pi R^2>1$, then 
$\cal O$ is relevant and the PBC fixed point is unstable.  
In this case the correction to $d_2-d_0$ becomes larger than the fixed 
point value, at sufficiently large $L$.  Since the state $|2>$ has 
$m=1$, and $m$ corresponds to a conserved quantum number at the PBC 
fixed point, the only non-zero matrix element of $\cos (\phi /R)$ in the
subspace of $|0>$ and $|2>$ is an off-diagonal one $<2|\cos (\phi
/R)|0>$.  Since this is larger than the diagonal matrix elements, the
energy difference, divided by $L$, scales as $L^y$, where $y$ is the 
renormalization group eigenvalue of the associated coupling constant, 
$y \equiv 1-x$.  On the other hand, for $4\pi R^2<1$, the PBC fixed 
point is stable, and in fact the corrections to the $d_i$'s from 
$\cal O$ vanish at large $L$. In this case, the state $|2>$ has $m=0$ 
(and $m_1^L$ or $m_1^R=1$).  Hence the first order corrections to the 
energies vanish and the leading corrections are second order, that is 
$d_2-d_0\propto L^{2y}$.  

At the OBC fixed point, the most relevant permitted operator is a 
product of the operators which change $n\pm 1$ from the two sides of 
the broken chain, that is $S^+_1S^-_L+\hbox{h.c.}$.  The $n=\pm 1$ 
operator has dimension $2\pi R^2(1\mp 2\delta/\pi)$.  Since the two 
sides of the broken chain are decoupled at the ideal OBC fixed point, 
the dimension of the product of operators on the two sides is simply 
additive, giving a total dimension for the leading operator of 
$4\pi R^2$.  Since there is no extra conserved quantum number at the 
OBC fixed point we expect this operator to have a non-vanishing 
expectation value in the states $|0>$ and $|2>$.  Hence the correction 
to $d_2-d_0$ should scale as $L^{y}$.

Exactly at $b=0$ or $1$, the set of allowed operators is reduced.  
For the PBC case, $b=1$, all local as well as $m\neq 0$ operators are 
forbidden by translation 
invariance.  The leading irrelevant operators correspond to
\begin{equation} 
\int dx (\partial\phi /\partial x)(\partial \phi /\partial x)^2\qquad 
\int dx (\partial\phi /\partial x)(\partial \phi /\partial t)^2, 
\end{equation}
of dimension 2.  [$\partial \phi /\partial x$ can be eliminated by a 
redefinition of $\phi$.]  
These dimension 2 operators are forbidden by symmetry for $M=0$ and so 
did not appear in \cite{A&E}. They contribute $1/L^2$ corrections to 
the energies.  For the $b=0$ case, since translational symmetry is 
broken, we can construct boundary operators but only at one side or 
the other of the broken link; we can not take a product of operators 
on both sides as above.  In addition, spin rotational invariance
forbids all operators with $n\neq 0$. Again $\partial \phi /\partial x$ 
is eliminated by a redefinition of $\phi$, so the leading irrelevant 
boundary operator is $(\partial \phi(0)  /\partial x)^2$ of dimension 
2\cite{A&E}.  This also contributes $1/L^2$ corrections to the energies.

When $b$ is close to 0 or 1, we obtain analytic $1/L^2$ finite-size 
corrections, as just discussed, with coefficients of $O(1)$, as well as
non-analytic finite size corrections, as previously shown, with
coefficients that vanish in the limit $b\to 0$ or 1.  

\section{Numerical studies on spectrum and renormalization}

We first consider the bulk quantities, $e_0$, $\mu$, $v_s$ and $R$, 
defined in Eq. (\ref{fss1}).  We determined these using 
both the DMRG and the Bethe ansatz technique~\cite{BetheAnsatz}.

These bulk quantities can be determined from the finite-size spectrum 
with PBC which can be derived from the Bethe ansatz~\cite{BAFS}. The 
structure of the spectrum up to $O(1/L)$ term agrees with the CFT 
prediction [$\delta=0, e_1 =0$ in Eq.~(\ref{fss1}) ].  By comparison, 
the various parameters are 
determined~\cite{BAFS} as:
\begin{eqnarray}
e_0 &=&\int_{-\Lambda}^{\Lambda}
			\epsilon_0(\eta) \sigma(\eta) d\eta ,\nonumber\\
  R &=&\frac{1}{\sqrt{4 \pi} \xi(\Lambda)} , \\ 
v_s &=&\frac{e}{2 \pi \sigma(\Lambda)}. \nonumber
\end{eqnarray}
Here the right-hand sides are given by
\begin{eqnarray}
\epsilon_0(\eta) 
      &=& \mu -\frac{\sin^2{\theta}}{\cosh{\eta}-\cos{\theta}},
				\nonumber\\
\xi(\eta) 
      &=& 1 - \frac{1}{2\pi}
         \int_{-\Lambda}^{\Lambda} K(\eta-\eta') \xi(\eta') d \eta' ,
				\nonumber\\
\sigma(\eta) 
      &=& \frac{1}{2 \pi}
          \left[ \frac{\cot{(\theta/2)}}{\cosh^2{(\eta/2)} 
                 + \cot^2{(\theta/2)} \sinh^2{(\eta/2)}} \right.\nonumber\\
      & & \mbox{} \left.
		- \int_{-\Lambda}^{\Lambda} 
		K(\eta - \eta') \sigma(\eta') d \eta' \right] , \\
\rho(\eta) 
      &=& \frac{1}{2 \pi} \left[
                 \frac{d K(\eta - \Lambda)}{d \eta}
                 - \int_{-\Lambda}^{\Lambda} K(\eta - \eta') 
			\rho(\eta') d \eta' \right], 
				\nonumber\\
e     &=& \frac{d \epsilon_0(\Lambda)}{d \Lambda}
          +\int_{-\Lambda}^{\Lambda}\epsilon_0(\eta)\rho(\eta) d\eta ,
				\nonumber
\end{eqnarray}
where $\cos{\theta}=a$ (see Eq.~(\ref{old1})) and the integral kernel 
is given by
\begin{equation}
  K(\eta) =
\frac{\tan{\theta}}{\tan^2{\theta}\cosh^2{(\eta/2)} + \sinh^2{(\eta/2)}} .
\end{equation}
The parameter $\Lambda$ is fixed from the filling factor $\nu=1/4$ by
\begin{equation}
  \int_{-\Lambda}^{\Lambda} \sigma(\eta) d \eta = \nu ,
\end{equation}
and the chemical potential $\mu$ is determined by the condition
$\epsilon_d(\Lambda)=0$, where
\begin{equation}
  \epsilon_d(\eta) = \epsilon_0(\eta) - \frac{1}{2 \pi}
     \int_{-\Lambda}^{\Lambda}K(\eta - \eta')\epsilon_d(\eta')d\eta' .
\end{equation}

The half-filled case $\nu = 1/2$ corresponds to $\Lambda=\infty$.
In this case the integral equations can be solved analytically by
Wiener-Hopf method~\cite{BAFS}, which gives
$R =  \sqrt{\frac{1 - \theta / \pi}{2 \pi}}$
and $v_s = \pi \sin{\theta}/(2 \theta)$, as deduced earlier from the
Bethe ansatz solution using other methods.\cite{desClois,Baxter}
In particular, for $a=\pm 0.5$ and $M=0$, the case considered in
Ref. \onlinecite{Q&F}, $v_s=3\sqrt{3}/4$, $3\sqrt{3}/8$, and 
$4\pi R^2=2/3$, $1/3$, respectively.  The Bethe ansatz results for 
$L(E_1-E_0)$, $e_2\equiv (E_2-E_0)/(E_1-E_0)$ and 
$e_3\equiv (E_3-E_0)/(E_1-E_0)$ are given in Table III.  Excellent 
agreement is obtained with the finite-size spectrum calculated using 
DMRG in \cite{Q&F}.

For the non-half-filled case, these equations cannot be solved
analytically. We solved them numerically by iterative numerical
integration. Since we have already taken the limit of large system size 
$L$ in analytic treatment, we can obtain highly accurate results with 
relatively small amount of computation.  Accuracy to six digits can be 
achieved within minutes on an IBM RS6000 workstation.  On the other 
hand, it would be difficult to calculate other quantities, for example 
orthogonality exponent, directly from Bethe Ansatz.

By fitting all four $E_i$'s obtained from DMRG according to
Eqs. (\ref{fss1}) and (\ref{fss2}), for the  cases we defined as 
{\sl ideal } PBC ($b=1$) and OBC ($b=0$), we find consistent bulk 
quantities, $e_0$, $\mu$, $v_s$, and $R$, which are listed in Table I. 
These values agree well with the Bethe Ansatz results, as also shown 
in Table I.  Thus the accuracy of our DMRG calculation is confirmed.  
We also determined, from DMRG, the non-zero forward scattering phase 
shift, $\delta$, for the case of OBC, as shown in Table I.

With these parameters, we have found that the finite-size corrections 
to the energies, $d_i$, found numerically agree with the low energy 
spectrum predicted by CFT for PBC and OBC and listed in Table II. The 
finite size values of $d_i$ are plotted versus $1/L$ in Figures 1 and 
2, and extrapolated at $L\to \infty$ by fitting with polynomials of 
$1/L$ .  The quality of these fittings confirms the prediction that 
only analytic finite size corrections exist in the cases with $b=0$ 
or 1.  

After having verified the low energy CFT spectrum, we confirm another 
prediction, namely that the orthogonality exponent between the ground 
states of two models described by the same Hamiltonian but different 
conformally invariant boundary conditions, is equal to the difference 
between their corresponding $d_0$'s.  The orthogonality exponent 
$\alpha$ is defined through the overlap between the two ground state 
wave functions, $|\phi\rangle$ and $|\phi_0\rangle$, for a chain of 
size $L$.  In particular, 
\begin{equation}
O(L)=\langle \phi |\phi_0\rangle\propto\left(\frac{1}{L}\right)^\alpha.
\label{old3}
\end{equation}
The overlap integrals $O(L)$ are evaluated 
numerically.  The $\alpha$ for $L\to\infty$ limit is obtained by 
extrapolating 
\begin{equation}
\alpha(L)= \displaystyle \frac{\ln O(L+4)- \ln O(L)}{\ln L - \ln (L+4)}.
\label{old4}
\end{equation}
We start by verifying the above prediction for the overlap between the 
ground states with {\sl ideal} PBC and OBC.  The {\sl ideal} OBC 
$d_0$'s for $L\to\infty$ limit (for the {\sl ideal} PBC, $d_0=0$) is 
obtained by extrapolating
\begin{equation}
d_0(L)=\frac{1}{12} -\frac{L(L+4)}{8\pi v_s}
         \left[E_0(L+4)-E_0(L)-4e_0\right].
\label{old5}
\end{equation}
In Fig. 3 we plot $\alpha(L)$, the exponent of the overlap between the 
{\sl ideal} OBC and PBC ground state wavefunctions, and the {\sl ideal} 
OBC $d_0(L)$, vs. $1/L$. Up to the two significant digits the 
extrapolated $\alpha$ and $d_0$ are equal for $L\to\infty$ limit, for 
both $a=0.5$ and $a=-0.5$.

We point out here that the low energy spectrum in Table II holds also
if the magnetization $M=0$, as it was shown in Ref.\onlinecite{Q&F}.  
For generic impurity strength, it was shown in that reference that 
the $M=0$ system flows either to OBC ($a>0$) or to PBC ($a<0$), by a 
detailed analysis of the size dependence of the low energy levels and 
by calculating the orthogonality exponent.  
Since the non-universal forward scattering 
phase shift $\delta=0$ for $M=0$, universality is recovered and all 
systems flowing to the OBC fixed point had an extrapolated $d_0=1/16$, 
in agreement with CFT prediction. Moreover, $1/16$ was also shown 
to be equal to the orthogonality exponent of the overlap between the 
ground states with PBC in the presence and in the absence of the 
impurity.  

Analogously, in the following, we will first show that the low energy 
spectrum for $M=1/4$ systems does flow to the PBC or OBC fixed points, 
depending on the sign of the interaction.  Then we will show the 
qualitative agreement between $\alpha(L)$ and $d_0(L)$ for arbitrary 
impurity strength $b$ at finite size.  

We start by showing how the low energy spectrum flows away from PBC 
towards the OBC fixed point for $a=0.5$ and vice versa for $a=-0.5$. 
We previously said that $\delta$ is not a universal property. This 
implies that, if a system flows to one of the two fixed points, it 
does not mean necessarily that the value of $\delta$ will be exactly 
the same as that of a chain with $b=0$ or $b=1$ (as given in Table I). 
Thus we expect $\delta$ to be some non-universal function of $b$ and 
$a$.  The finite size spectrum should be given by the OBC formula of 
Eq. (\ref{fss2}) and Table II for $a=0.5$ and all $b$ but with some 
unknown function $\delta (b)$.  Likewise, for $a=-0.5$ the spectrum 
should be given by the PBC formula of Eq. (\ref{fss1}) and Table II 
for all $b$ but with some other unknown function $\delta (b)$.  
$\delta (b)$ should approach zero as $b\to 1$ for $a=-0.5$ and 
$\delta (b)$ should approach its OBC value ($\approx -0.29\pi$) as 
$b\to 0$ for $a=0.5$ since in these two cases $1-b$ or $b$ produce no 
relevant operators and the marginal operator has a coupling constant 
which goes to zero.  

For the four energy levels we decided to  calculate numerically, 
$E_3-E_1=E_2-E_0$ always holds, as seen from Table II.  Notice that 
the forward scattering phase shift $\delta$  generated by the impurity 
contributes to $E_1-E_0$ (and also  $E_3-E_2$).  In order to simplify 
the analysis, we will concentrate only on $E_2-E_0$ ($i.e.$ $d_2-d_0$), 
which instead depends on the universal properties identifying each 
fixed point.  Let us denote $\Delta(a,b)= d_2(a,b)-d_0(a,b)$.  Then, 
for small deviations of $b$ away from one of the two fixed point values
$b_*=1,0$, the following scaling should hold
\begin{equation}
\Delta(a,b)-\Delta(a,b_*)\propto L^{\gamma(a,b_*)}.
\label{old6}
\end{equation}
These exponents can be obtained by calculating the first non-vanishing 
correction to $\Delta(a,b)$, within a perturbation expansion in
$b-b_*$, following the prescriptions given in the previous section.  
The predicted $\gamma$'s are listed in Table IV.  In Fig.4, 
for $a=\pm 0.5$, we plot $\ln (\Delta(a,b)-\Delta(a,b_*)) / \ln L$ vs. 
$1/\ln L$.  The extrapolated scaling exponents $\gamma$ in the figure 
agree with those listed in Table IV, up to the two significant digits. 
These results have a very simple interpretation.  If $\gamma <0$, the 
deviation from the fixed point is not a relevant perturbation and the 
system flows back to the $b_*$-fixed point, which is therefore stable.  
On the contrary, if $\gamma >0$, the perturbative correction blows up 
as $L\to\infty$, which implies that the system flows away from the 
$b_*$ fixed point, which is therefore unstable.  Again, our results 
demonstrate that, for repulsive interactions, the stable fixed point 
boundary condition is the OBC, and vice versa for attractive 
interaction.

We have also calculated $\Delta(a,b)$ for arbitrary deviations $b-b_*$, 
namely for $b=0.05,0.15,\ldots,0.85,0.95$.  To analyze the data, we 
made a very simple scaling ansatz in order just to reproduce the 
correct scaling behavior close to the OBC fixed point.  Specifically, 
$\Delta(a,b)$ is plotted in Fig.5 vs. 
$\arctan \left[L^{-0.2}\tan(b\frac{\pi}{2})\right] / \pi$
for $a=0.5$, and in Fig.6 vs. 
$\arctan \left[ L^{0.28} \tan (b\frac{\pi}{2})\right] / \pi$
for $a=-0.5$.  The arrows in the figures show towards which fixed point 
the system flows as the chain length $L$ increases.  We see 
qualitatively how the excitation energy flows between OBC and PBC fixed 
points.  The proposed scaling is 
$\tan \left[(\Delta - \frac 1 2) \pi \right]$
$\approx$ $L^\gamma \tan (b \frac \pi 2)$.
Surprisingly, our scaling ansatz, which is in principle valid only 
close to $b=0$ and $L^\gamma$ being small, seems to give a good 
description of the data for a wide range of $b$'s.  
For $\frac b 2 L^\gamma \to 0$, we have the scaling relation:
\begin{equation}
\Delta - \frac 1 2  \to   \frac b 2 L^\gamma .
\end{equation}

Finally we compare the orthogonality exponent $\alpha(L)$ of the 
overlap between the ground state with PBC and in the presence of the 
impurity, and the ground state with PBC but in the absence of the 
impurity.  According to CFT, this exponent should also be equal to 
$d_0$ for chain with impurity. We  plot $\alpha(L)$ and $d_0(L)$ vs. 
the same variables as used above for $\Delta$.  The resulting data are 
plotted in Fig.7 for $a=0.5$, and in Fig.8 for $a=-0.5$.  They show 
that $\alpha(L)$ and $d_0(L)$ are of similar magnitude  and have 
similar flows for various chain lengths and impurity strengths.  Again, 
an important remark is needed regarding the forward scattering phase 
shift.  Since we can not exclude its generation at the fixed point, 
this would imply, for instance, that the orthogonality exponent for 
attractive interaction is, in general, non-zero.  With long enough 
chains it should be possible to determine $\delta (b)$  and check that 
both $\alpha (L)$ and $d_0(L)$ extrapolate to the same numbers, 
determined by these phase shifts.  However, with the limited lengths 
available ($L\leq 60$) we have not found it possible to do this.  This 
is presumably due to the unknown value of the phase shift, the slow 
convergence of the finite-size corrections due to the small exponent 
in the non-analytic corrections $(L^{-0.2}$ for $a=0.5$) and the 
relatively large coefficient in the analytic corrections.  The 
situation appeared somewhat better in the $M=0$ case \cite{Q&F}.

\section{Discussion on boundary condition}
With periodic boundary conditions, the boson may be written in terms of 
decoupled left and right-moving parts:
\begin{equation} \phi (t,x)=\phi_L(t+x)+\phi_R(t-x).\end{equation}
The dual field is defined as $\tilde \phi \equiv \phi_L-\phi_R$.
The boundary condition, $\phi (0^\pm)=\pm \phi_0$ couples  left
movers to right movers, separately on both sides of the origin:
\begin{eqnarray}\phi_L(t,0^-)&=&-\phi_R(t,0^-)-\phi_0\nonumber \\
\phi_R(t,0^+)&=&-\phi_L(t,0^+)+\phi_0.\label{bc}\end{eqnarray}
On the other hand the fields on opposite sides of the origin are not
coupled by the boundary condition. This means that an incoming
excitation at $x<0$ is reflected, with unit probability with a phase
shift proportional to $\phi_0$ and similarly for an incoming excitation 
at $x>0$.  This boundary condition is the same as the one used in Ref. 
\onlinecite{A&L} where a different basis was used.

This coupling of left and right movers by relevant backscattering 
changes Green's functions.\cite{A&E} To illustrate this, we consider 
the $M=0$ case where there is no forward scattering term, $\phi_0=0$.  
One term in the bosonized representation of $S^-_j$ is:
\begin{equation} 
\begin{array}{lcl}
S_j^- & \propto  &e^{i2\pi R\tilde \phi +i\phi /R}+\ldots \\
      & \propto  &e^{i(2\pi R+1/R)\phi_L}e^{i(-2\pi R+1/R)\phi_R}+\ldots
\end{array}
\label{bos}
\end{equation}
The Green's function, with PBC  factorizes into separate Green's 
functions for left and right movers, giving:
\begin{equation}
\begin{array}{lcl}
\langle S^+(t,x)S^-(0,x') \rangle 
	&\!\!\!\!\! \propto \!\!\!\!\! & 
                 (t+x-x')^{-(2\pi R+1/R)^2/4\pi} \\
        &  &\times (t-x+x')^{-(-2\pi R+1/R)^2/4\pi}+\ldots .
\end{array}
\end{equation}
On the other hand, with the perfectly reflecting boundary condition,
this correlation function equals 0  if $x$ and $x'$ are on opposite 
sides of the origin.  To calculate the Green's function when they are 
on the same side (say $x,x'>0$) we use a standard device from boundary 
CFT.\cite{Cardy3}  We can regard the second boundary condition of Eq. 
(\ref{bc}) as defining $\phi_R$ for all $x>0$ as the analytic 
continuation of $\phi_L$ to the negative $x$-axis: 
\begin{equation} \phi_L(t,-x)\equiv -\phi_R(t,x)+\phi_0. \end{equation}
The reason this works is because the boundary condition is true at all 
$t$ and $\phi_L$ and $\phi_R$ only depend on the variables $t+x$ and 
$t-x$, respectively.  We emphasize that this analytically continued 
$\phi_L$ is {\it not}  the original $\phi_L$ at $x<0$ which is 
completely decorrelated from the fields at $x>0$.  Using this approach,
$S^-(t,x)$ becomes bilocal: \begin{equation} S^-(t,x)\propto e^{i(2\pi
R+1/R)\phi_L(t+x) -i(-2\pi R+1/R)\phi_L(t-x)}+\ldots .\end{equation}
Instead of getting a product of 2-point Green's functions for left and
right moving fields, we now obtain a 4-point Green's function for
left-movers only, giving:
\begin{equation} 
\begin{array}{lcl}
\langle S^+(t,x)S^-(0,x') \rangle & \propto &
	\left|{xx'\over t^2-(x+x')^2}\right|^{-1/4\pi R^2+\pi R^2}\\
  & & \times \left|{1\over t^2-(x-x')^2}\right|^{1/4\pi R^2+\pi R^2} \\
  & & \times \left|{t-x+x'\over t+x-x'}\right| +\ldots 
\end{array}
\label{4pt}
\end{equation} 
In the limit, $x,x'\to 0^+$, 
\begin{equation} S^-(t,0)\propto e^{i2\pi R\tilde \phi (t,0^+)}
\propto  e^{i4\pi R\phi_L(t,0^+)},\end{equation}
giving:\begin{equation}
<S^+(t,0^+)S^-(0,0^+)>\propto |t|^{-4\pi R^2}+\ldots ,\end{equation}
which is the limiting behaviour of Eq. (\ref{4pt}) (with  $xx'$ set 
equal to a value of order of the short distance cutoff, i.e. the 
lattice spacing). Note that without pinning :
\begin{equation}
\begin{array}{l}
\langle e^{i2\pi R\tilde \phi (t,0)}
        e^{-i2\pi R\tilde \phi (0,0)} \rangle \\
= \langle e^{i2\pi R\phi_L (t,0)}
          e^{-i2\pi R\phi_L (0,0)}\rangle 
   \langle e^{-i2\pi R\phi_R (t,0)}
           e^{i2\pi R\phi_R (0,0)}\rangle  \\
\propto |t|^{-2\pi R^2}. 
\end{array}
\end{equation}
Thus pinning increases the exponent  by a factor of 2.  Perhaps 
surprisingly, pinning $\phi (0)$ changes correlation functions 
involving $\tilde \phi$.  This effect ultimately arises because $\phi$ 
and $\tilde \phi$ cannot be regarded as being independent.  
$\partial \tilde \phi /\partial x$ is canonically conjugate to $\phi$.  
Therefore when $\phi$ gets pinned we must take into account the effect 
on $\tilde \phi$.  Different results were obtained in Ref. 
\onlinecite{Fink} because this effect was not taken into account.  In 
particular, a factor of two discrepancy in the exponent occurring in  
the electron Green's function (at $x=x'=0$) arose from the mechanism 
explained here.  An alternative way of understanding the
disagreement with Ref. \cite{Fink} can be found in Ref. \cite{fabr}
 and in Ref. \cite{furu}.

\section{Summary}

In summary, this paper studies the effects of an impurity in a 
non-half-filled 1D Luttinger liquid.  Our numerical study for the low 
energy spectrum confirms previous results.  That is, we show that the 
low energy spectrum of a chain with an impurity flows to the spectrum 
of a chain with PBC if the interaction is attractive, and to the 
spectrum of a chain with OBC if the interaction is repulsive.
The behavior of a non-half-filled system is shown to be the same as 
that of a half-filled one and the presence of particle-hole symmetry
is not essential for this behaviour.  
Therefore, we numerically confirm that the properties of a generic 1D 
Luttinger liquid with an impurity are indeed those predicted by Kane 
and Fisher\cite{K&F}.

  \acknowledgements

The DMRG calculation was completed on IBM RISC-6000 at International
Center for Theoretical Physics, Trieste.  Mobility in Europe involved 
in this research project was partly sponsored by EEC under contract ERB 
CHR XCT 940438. One of us (MF) acknowledge partial support by INFM, 
project HTSC.  The research of MO and IA was supported in part by NSERC 
of Canada.

\begin{table}[bt]
\caption{
Bulk quantities: site energy $e_0$, chemical potential $\mu$, spin 
velocity $v_s$, boson radius $R$, and forward scattering phase shift 
for boundary conditions $b=0,1$: $\delta$, for quarter filling $S=1/2$ 
XXZ chains with $a=0.5$ and $a=-0.5$.  Values are obtained 
independently from DMRG (RG) and Bethe Ansatz (BA) calculations.
}
$$
\begin{array}{ccccc}
a       &  0.5 (RG) & 0.5 (BA)   & -0.5 (RG) & -0.5 (BA) \\
e_0     & -0.220487   & -0.220487 & -0.233179  & -0.233179 \\
\mu     &  -1.1335    & -1.13349 & -0.32120    & -0.321201 \\
v_s     &  0.833      & 0.832900 & 0.500       &  0.500854 \\
4\pi R^2&  1.20       & 1.18383  & 0.72        &  0.717842 \\
\frac{\delta}{\pi} \mbox{ } (b=1)&  0  & 0 & 0    & 0\\
\frac{\delta}{\pi} \mbox{ } (b=0)&  -0.29 & N/A  & -0.19 & N/A
\end{array}
$$
\end{table}

\begin{table}[hbt]
\caption{
Conformal field theory prediction of $d_i$ for PBC and OBC fixed points 
(see definition in Eq. (\protect\ref{fss}) and predictions in Eqs. 
(\protect\ref{fss1}, \protect\ref{fss2})). $\delta$ is the forward 
scattering  phase shift.  The expressions for PBC are different for 
$a=0.5$ and $a=-0.5$ . 
}
$$
\begin{array}{cccc}
a     &(\frac{1}{2},PBC)             
         &(-\frac{1}{2},PBC)
            & (\pm\frac{1}{2},OBC) \\
 d_0  &\pi R^2 (\frac{\delta}{\pi})^2 
         &\pi R^2 (\frac{\delta}{\pi})^2
            & \frac{1}{16}\!\!+\!\! \pi R^2(\frac{\delta}{\pi})^2\\
 d_1  &\pi R^2 (1\!\!-\!\!\frac{\delta}{\pi})^2 
         &\pi R^2 (1\!\!-\!\!\frac{\delta}{\pi})^2
            & \frac{1}{16}\!\!+\!\! \pi R^2(1\!\!-\!\!\frac{\delta}{\pi})^2\\
 d_2  &\frac{1}{4\pi R^2}\!\!+\!\!\pi R^2 (\frac{\delta}{\pi})^2 
         &1\!\!+\!\!\pi R^2 (\frac{\delta}{\pi})^2
            & \frac{9}{16}\!\!+\!\! \pi R^2(\frac{\delta}{\pi})^2\\
 d_3  &\frac{1}{4\pi R^2}\!\!+\!\!\pi R^2 (1\!\!-\!\!\frac{\delta}{\pi})^2 
         &1\!\!+\!\!\pi R^2 (1\!\!-\!\!\frac{\delta}{\pi})^2
            & \frac{9}{16}\!\!+\!\! \pi R^2(1\!\!-\!\!\frac{\delta}{\pi})^2\\
\end{array}
$$
\end{table}

\begin{table}[bt]
\caption{Properties of $M=0$ model.}
$$ \begin{array}{ccc}
a & .5 & -.5 \\
L(E_1-E_0)&	\pi \sqrt{3}/2	&\pi \sqrt{3}/8 \\
e_2 (b=0)	&3/2 & 3\\
e_2(b=1)	&9/4&  6\\
e_3(b=0)	&5/2& 4\\
e_3(b=1) &13/4 & 7\\
\end{array}$$
\end{table}

\begin{table}[bt]
\caption{
The scaling exponent $\gamma$ for small deviation from fixed points 
$(a,b_*)$.  $\gamma>0$ and $\gamma<0$ means the energy spectrum flows 
away from and flows to the fixed points, respectively.
}
$$
\begin{array}{ccccc}
(a,b_*)&(\frac{1}{2},1)&(\frac{1}{2},0)&(-\frac{1}{2},1)&(-\frac{1}{2},0)\\
\gamma& 1-1/4\pi R^2 &1-4\pi R^2     &  2(1-\frac{1}{4\pi R^2})&1-4\pi R^2\\
      &    0.17      &-0.20          &  -0.78    &0.28\\
\end{array}
$$
\end{table}

\begin{figure}[hat]
\epsfxsize=\columnwidth\epsfbox{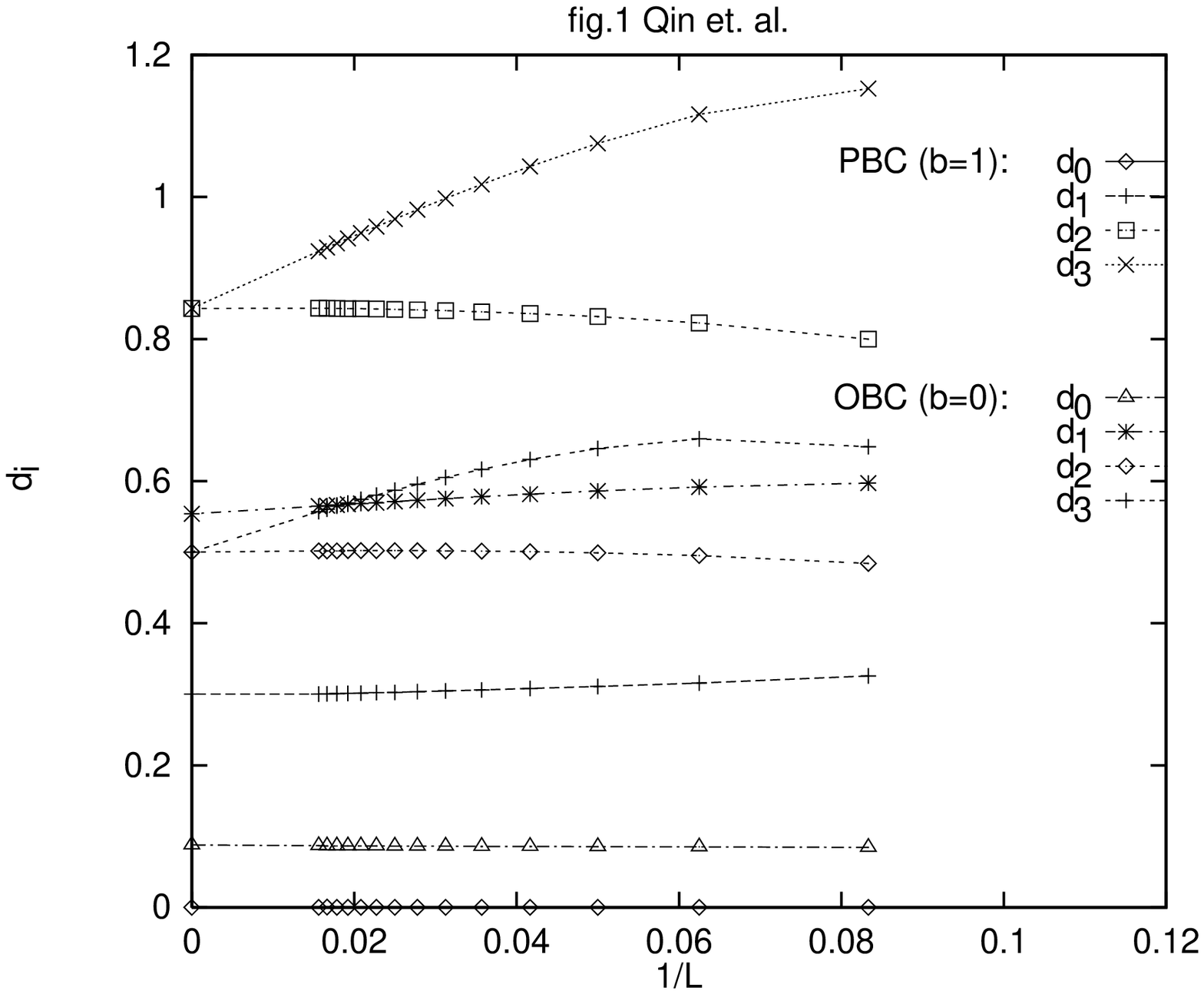}
\caption{
For $a=0.5$, $b=0$ and $b=1$, the $d_i$ defined in 
Eq.(\protect\ref{fss}\protect) is 
plotted for chain lengths $L=8,16,\ldots,64$ as a function of $1/L$. We 
extrapolated the $d_i$ of chain lengths $L=32$ to $52$ to $L\to\infty$ 
limit by polynomials of $1/L$ to the second order. The extrapolated 
values for $d_i$ are in agreement with the conformal field theory 
predictions given in Table.II, with $\delta=0$ for $b=1$ and 
$\delta=-0.29$ for $b=0$.
}
\end{figure}

\begin{figure}[hat]
\epsfxsize=\columnwidth\epsfbox{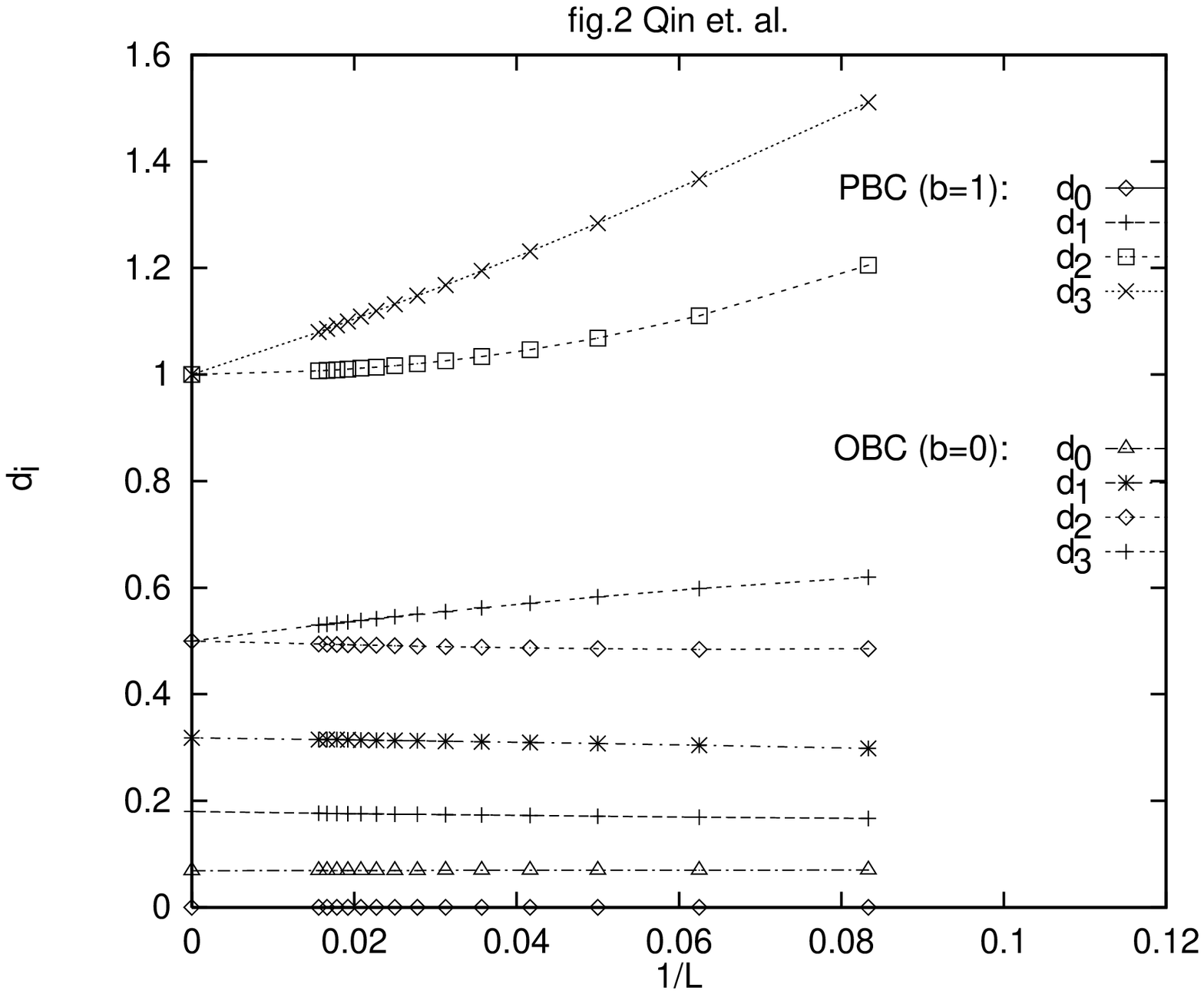}
\caption{
For $a=-0.5$, $b=0$ and $b=1$, the $d_i$ defined in 
Eq.(\protect\ref{fss}\protect) is 
plotted for chain lengths $L=8,16,\ldots,64$ as a function of $1/L$. We 
extrapolated the $d_i$ by data of chain lengths $L=32$ to $52$ to 
$L\to\infty$ limit by polynomials of $1/L$ to the second order.  The 
extrapolated values for $d_i$ are in agreement with the conformal field 
theory predictions given in Table.II, with $\delta=0$ for $b=1$ and 
$\delta=-0.19$ for $b=0$.
}
\end{figure}

\begin{figure}[hat]
 \epsfxsize=\columnwidth\epsfbox{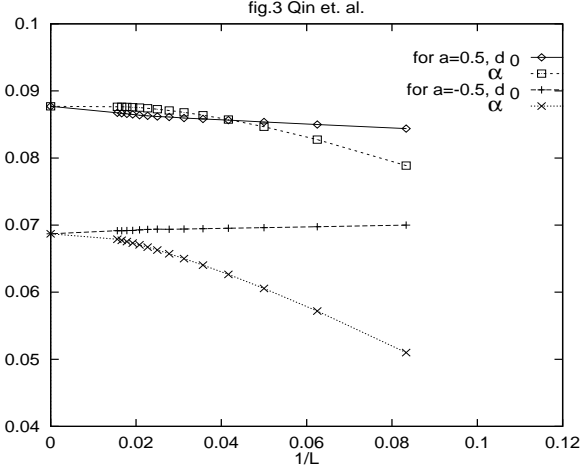}
\caption{
The orthogonality exponent $\alpha(L)$ and ground state scaling 
dimension $d_0(L)$ for OBC chains, (see Eq.(\protect\ref{old4}) and Eq. 
(\protect\ref{old5})), are extrapolated to $L\to\infty$ limit by 
polynomials of $1/L$ to the second order. We show the extrapolated 
values for $\alpha$ and $d_0$ are equal as conformal field theory 
predicts.
}
\end{figure}

\begin{figure}[hbt]
 \epsfxsize=\columnwidth\epsfbox{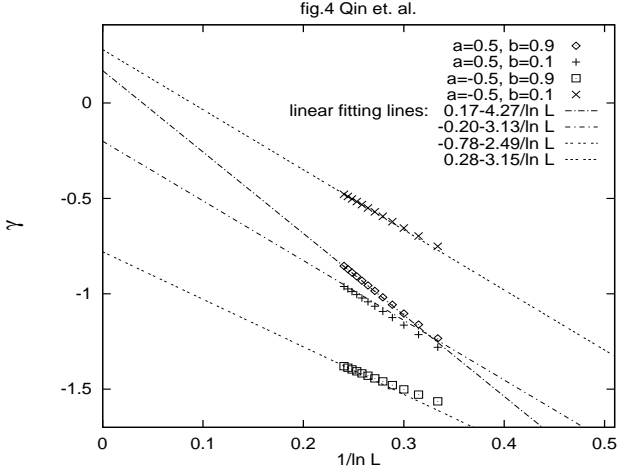}
\caption{
The scaling exponent $\gamma$ for excitations energy near fixed points, 
see Eq. (\protect\ref{old6}), is given for chains with lengths around 
50.  The calculated $\gamma$ are in agreement with the theoretical 
values listed in Table IV.
}
\end{figure}
 
\begin{figure}[hbt]
 \epsfxsize=\columnwidth\epsfbox{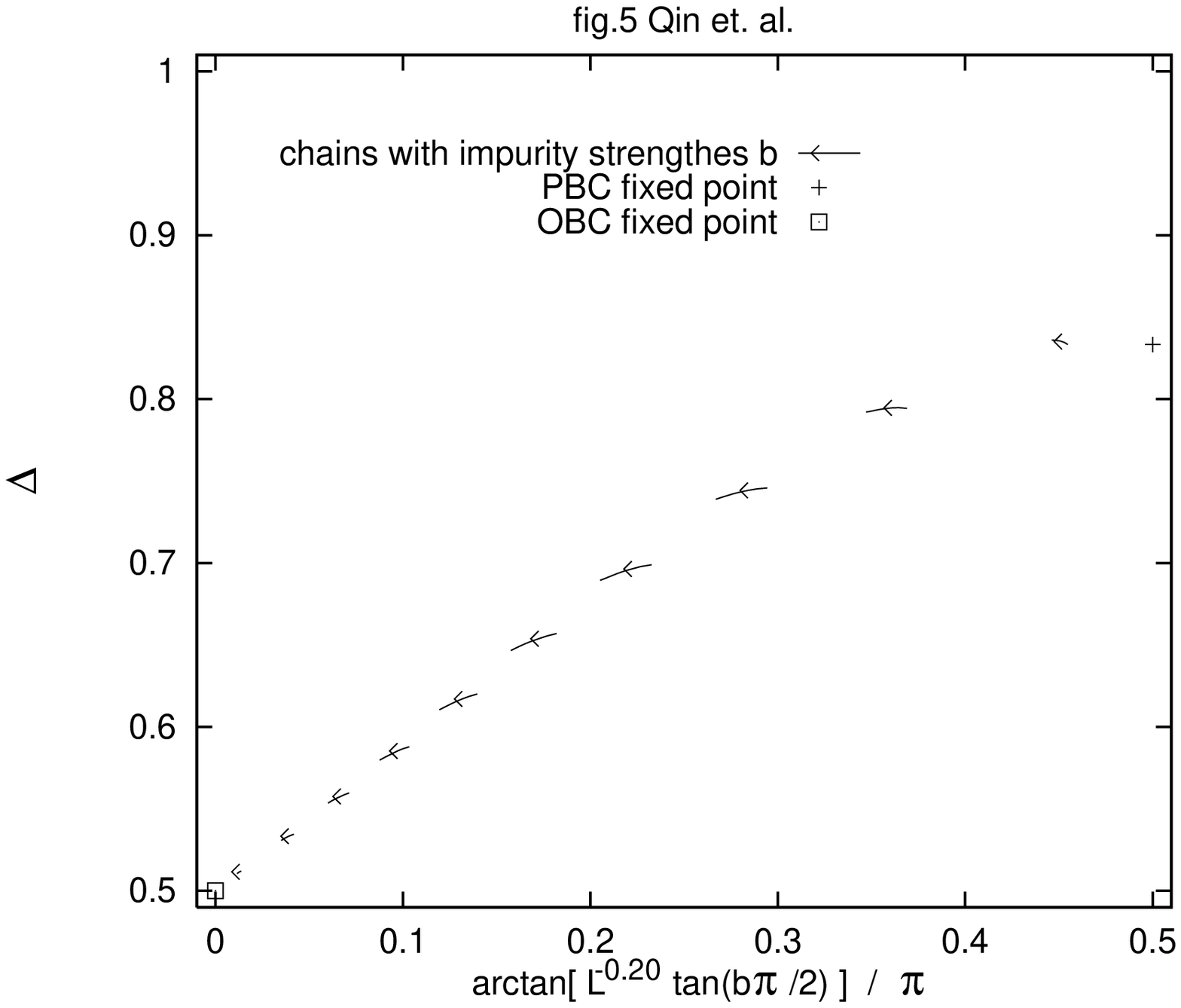}
\caption{
The renormalization group flow of the excitation energy to OBC ($b=0$) 
fixed point for $a=0.5$ is qualitatively shown in this figure for the
range $L=20$ to $50$.  The line segments from left to right of the plot 
are for $b=0.05,0.15,...,0.95$, respectively, and the arrow indicates 
how excitation energy $\Delta(0.5,b)$ flows when $L$ increases.  
}
\end{figure}
 
\begin{figure}[hbt]
 \epsfxsize=\columnwidth\epsfbox{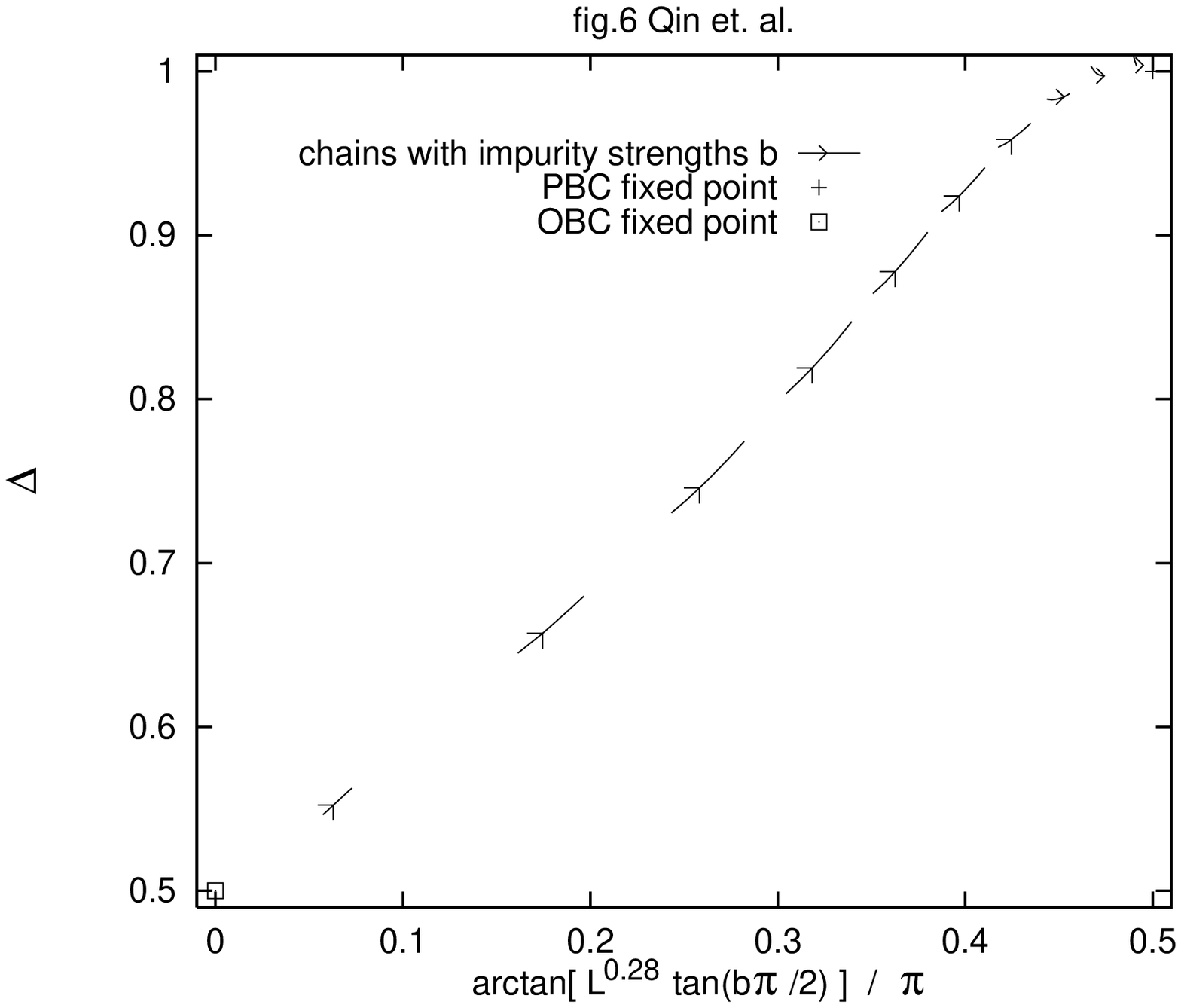}
\caption{
The renormalization group flow of the excitation energy to PBC ($b=1$) 
fixed point for $a=-0.5$ is qualitatively shown in this figure for the
range $L=20$ to $50$. The line segments from left to right of the plot 
are for $b=0.05,0.15,...,0.95$, respectively, and the arrow indicates 
how excitation energy $\Delta(-0.5,b)$ flows when $L$ increases.  
}
\end{figure}

\begin{figure}[hbt]
 \epsfxsize=\columnwidth\epsfbox{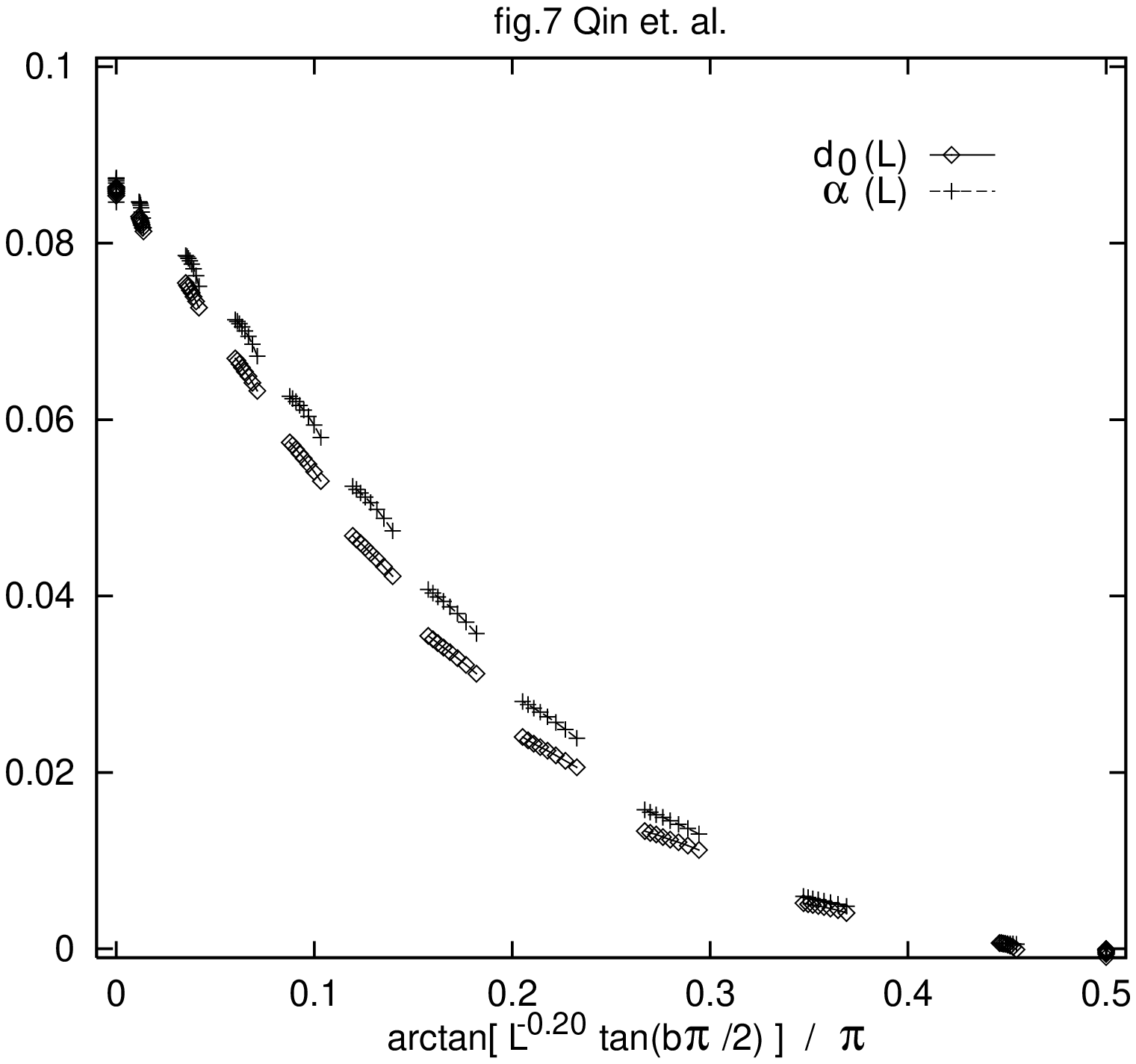}
\caption{
The orthogonality exponent $\alpha(L)$ and ground state scaling 
dimension $d_0(L)$ for impurity chains with $a=0.5$, (see Eqs. 
(\protect\ref{old4}) and Eq.(\protect\ref{old5})), are plotted in this 
figure.  The line segments from left to right of the plot are for 
$b=0.0,0.05,0.15,...,0.95,1.0$, respectively. The figure shows 
qualitatively how $\alpha(L)$  and $d_0(L)$ flow to OBC fixed point as 
the chain length increases.  
}
\end{figure}
 
\begin{figure}[hbt]
 \epsfxsize=\columnwidth\epsfbox{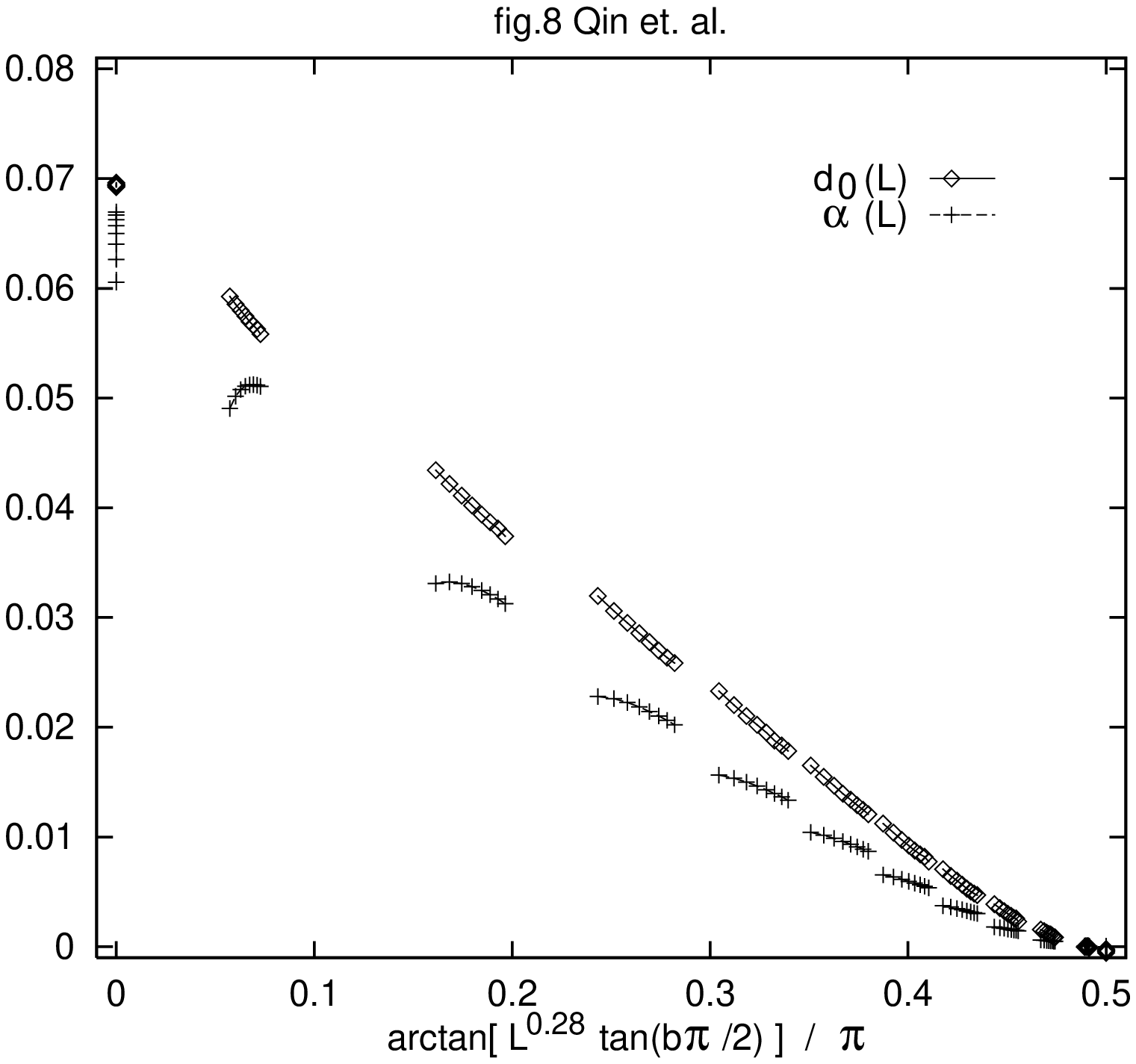}
\caption{
The orthogonality exponent $\alpha(L)$ and ground state scaling 
dimension $d_0(L)$ for impurity chains with $a=-0.5$, (see Eqs. 
(\protect\ref{old4}) and Eq. (\protect\ref{old5})), are plotted in this 
figure.  The line segments from left to right of the plot are for 
$b=0.0,0.05,0.15,...,0.95,1.0$, respectively. The figure shows 
qualitatively how $\alpha(L)$  and $d_0(L)$ flow to OBC fixed point as 
the chain length increases.  
}
\end{figure}
\end{document}